\documentclass[twocolumn,final]{svjour3} % onecolumn (standard format)
\smartqed  % flush right qed marks, e.g. at end of proof

\usepackage[export]{adjustbox}
\usepackage{blkarray, bigstrut}
\usepackage{xparse}

\usepackage{enumitem}
\usepackage{multirow} 
\usepackage{algorithmicx}
\usepackage{algpseudocode}
\usepackage{algorithm}
\usepackage{amsmath}
\usepackage{rotating}
\usepackage{tabularx}

\usepackage{mathtools}
\usepackage{amsmath}

\usepackage{kantlipsum} %with the next two command (commath, allowdisplaybreaks) -> allow to break the formulas along the pages
\usepackage{commath}
\allowdisplaybreaks
\usepackage{mathtools}  %with the next command 
\usepackage{verbatim}
\usepackage{xr-hyper} 
\usepackage{enumitem}
\usepackage{multirow}
\usepackage{hhline}
\usepackage[table,xcdraw]{xcolor}
 \usepackage{graphicx,booktabs}
 \usepackage{longtable}
 \usepackage{subcaption}

\usepackage[hyphens]{url}
\usepackage{hyperref}
\hypersetup{breaklinks=true}
\usepackage{dblfloatfix}

\usepackage{epstopdf}
\usepackage{wrapfig}
\usepackage{latexsym}
\usepackage{amssymb}
\usepackage{amsfonts}
\usepackage{amsmath} %[cmex10]
\usepackage{graphicx}
\usepackage{latexsym}
\usepackage{booktabs}
\usepackage[style=base]{caption}
\usepackage{subcaption} %******************* This package has conflict with sufig and subfigure
\usepackage{breqn}

\definecolor{white}{rgb}{0.88,1,1}
\algnewcommand\algorithmicinput{\textbf{INPUT:}}
\algnewcommand\INPUT{\item[\algorithmicinput]}
\algnewcommand\algorithmicoutput{\textbf{OUTPUT:}}
\algnewcommand\OUTPUT{\item[\algorithmicoutput]}
\algdef{SE}[DOWHILE]{Do}{doWhile}{\algorithmicdo}[1]{\algorithmicwhile\ #1}

\makeatletter
\newcounter{algorithmbis}
\renewcommand{\thealgorithmbis}{\arabic{algorithmbis}}
\def\algorithmbis{\@ifnextchar[{\@algorithmbisa}{\@algorithmbisb}}
\def\@algorithmbisa[#1]{%
  \refstepcounter{algorithmbis}
  \trivlist
  \leftmargin\z@
  \itemindent\z@
  \labelsep\z@
  \item[\parbox{0.49\textwidth}{%
    \hrule
    \noindent\strut\textbf{Algorithm \thealgorithmbis} #1
    \hrule
  }]\hfil\vskip+0em%
}
\def\@algorithmbisb{\@algorithmbisa[]}

\makeatother

\usepackage[table]{xcolor}

\usepackage{tikz}
\usetikzlibrary{fit} 

\usepackage{color}

\colorlet{BLUE}{blue}
\usepackage{colortbl}
\definecolor{white}{RGB}{155, 227, 247}

\newif\ifcommentson

\newif\ifextended
\newif\ifshortver

\shortvertrue
\newcommand{\optional}[1]{\ignorespaces}

\newif\ifrevisionactive
\newif\ifshowdeleted
\revisionactivetrue
\allowdisplaybreaks

\definecolor{maroon}{cmyk}{0,0.87,0.68,0.32}

\begin{document}

\title{IE-RAP: An Intelligence and Efficient Reader\\ Anti-Collision Protocol for Dense RFID Networks}

\titlerunning{IE-RAP: An Intelligence and Efficient Reader\\ Anti-Collision Protocol for Dense RFID Networks}

	\author{Hadiseh Rezaei \and Rahim Taheri \and Mohammad Shojafar}

\institute
    {H. Rezaei and R. Taheri \at PAIDS Research Centre, School of Computing, University of Portsmouth, UK \email{(hadiseh.rezaei,rahim.taheri)@port.ac.uk}
	\and
	M. Shojafar \at 5G/6G Innovation Centre, ICS, University of Surrey, UK
		\email{(m.shojafar)@surrey.ac.uk}}
 
\date{Received: .... / Revised: ..../ Accepted: ....}
\maketitle

\begin{abstract}
An advanced technology known as a radio frequency identification (RFID) system enables seamless wireless communication between tags and readers. This system operates in what is referred to as a dense reader environment, where readers are placed close to each other to optimize coverage. However, this setup comes with its challenges, as it increases the likelihood of collisions between readers and tags (reader-to-reader and reader-to-tag), leading to reduced network performance. To address this issue, various protocols have been proposed, with centralized solutions emerging as promising options due to their ability to deliver higher throughput. In this paper, we propose the Intelligence and Efficient Reader Anti-collision Protocol (IE-RAP) that improves network performance such as throughput, average waiting time, and energy consumption, which employs a powerful combination of Time Division Multiple Access (TDMA) and Frequency Division Multiple Access (FDMA) mechanisms. IE-RAP improves the efficiency of RFID networks through techniques such as the SIFT function and distance calculation between readers. By preventing re-read tags and ensuring the on-time release of the communication channel, we effectively eliminate unnecessary collisions. Our simulations emphasize the superiority of our proposed method, it increases 26\% throughput, reduces 74\% the average waiting time, and lower by 52\% the energy consumption compared to existing approaches. Importantly, our solution supports the seamless integration of mobile readers within the network.

\keywords{ Radio Frequency Identification (RFID)\and Reader Collisions \and Throughput \and Energy Consumption \and Average Waiting Time}

\end{abstract}
% \end{frontmatter}

\section{Introduction}\label{sez:1}
A radio frequency identification (RFID) system operates using near/far-field wireless communication to identify tagged objects. This system boasts several advantages, including rapid identification, a long transmission range, and efficient data storage, making it applicable in diverse fields such as supply chain management, object or animal tracking, medical care, and so on~\cite{haibi2022systematic,du2025vehealth}. Comprising two core components - tags and readers - RFID systems can be integrated with wireless sensor networks, offering promising platforms for wireless rechargeable sensor networks, as seen in various researches~\cite{rezaie2023shared,akbari2025application}.  

The tags are small electrical components physically attached to objects to carry essential information. They can be categorized into three types: active, semi-active, and passive. Passive tags, the most commonly used due to their affordability, do not require an explicit power supply; instead, they draw energy from readers' wave transmissions to respond directly. On the other hand, active tags have their own power supply and perform all functions using this internal power supply. Semi-active tags fall between these two categories, utilizing internal power for certain tasks and reader-transmitted energy for response. In most RFID systems, passive tags prevail due to their cost-effectiveness~\cite{zeng2024fast,el2024introducing,rezaei2025lightchain}.

The readers are small electrical components that scan their surroundings and identify tags through wave transmissions. Each reader has two key ranges: the read range and the interference range. The read range denotes the maximum distance at which a reader can read nearby tags, while the interference range indicates the distance from which a reader's emitted waves interfere with those of other readers in the vicinity. It is worth noting that the maximum distance between these two ranges is influenced by the reader's transmission power~\cite{mbacke2018survey,akdag2022novel}. This structure enhances traditional RFID tags' sensing and computation capabilities, with a particular focus on reducing charging delays~\cite{fu2015optimal,cheng2023enhanced}.
\begin{figure}[!htbp]
\centering 
\includegraphics[width=0.5\columnwidth]{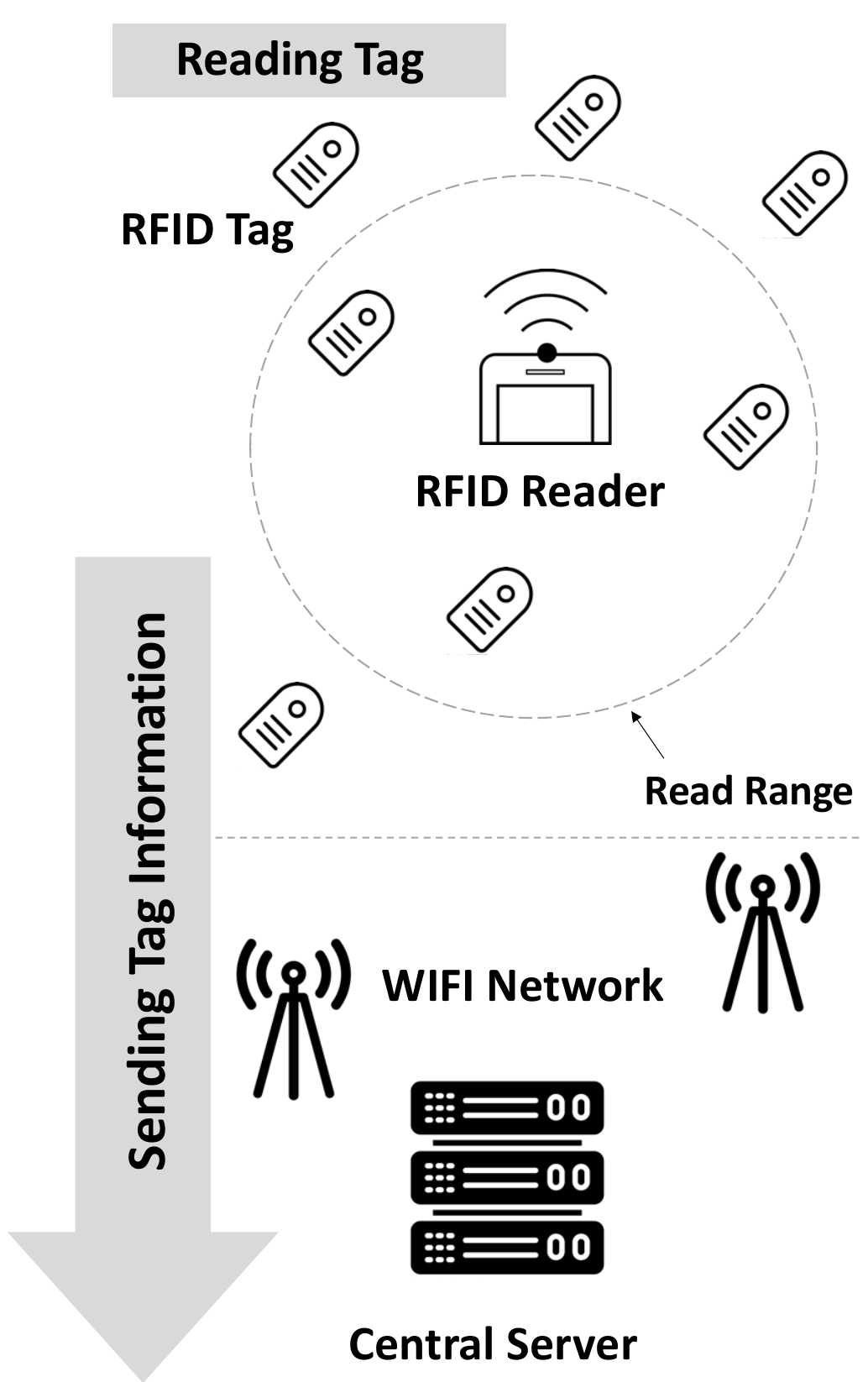}
\caption{ \small The RFID network structure}
\label{fig:RFIDnetworkstructure}
\end{figure}

As demonstrated in Figure~\ref{fig:RFIDnetworkstructure}, the reader-to-tag identification protocol enables readers to identify nearby tags and read their information. This capability is essential due to the mobility of readers, because their neighbors and their surrounding tags will be changed. The collected data is then stored and transmitted to a central server via wireless or wired communication. In certain scenarios, a single reader may not be sufficient to cover the entire target area. The RFID networks need multiple readers, creating what is referred to as a dense reader environment (DRE)~\cite{percy2025strategies,ai2022anti,rezaie2016sift}.

\begin{figure}[!htbp]
\centering 
\includegraphics[width=1\columnwidth]{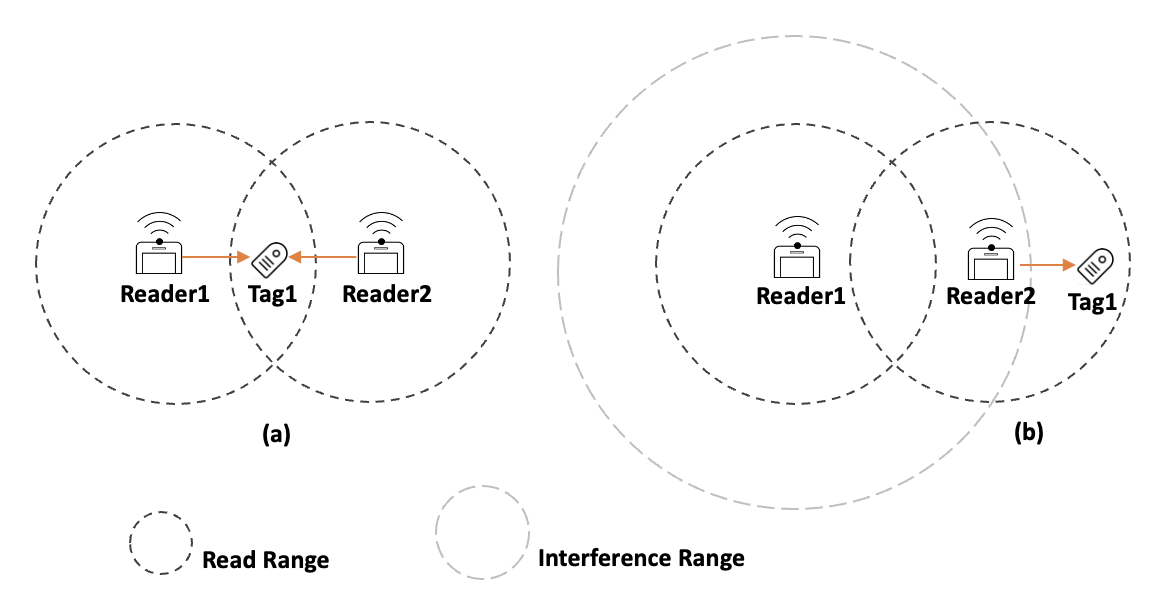}
\caption{ \small (a) Reader-to-Tag Collision and (b) Reader-to-Reader Collision}
\label{fig:reader-collision}
\end{figure}

In a DRE, readers are positioned closely together to effectively cover the environment. However, this proximity introduces the risk of collisions between readers and tags, leading to a decrease in the system's performance. Therefore, finding solutions for collision problems is one of the major focuses of RFID systems. These collisions typically occur at the Mac layer, where simultaneous transmission between readers causes issues. In dense reader environments, the following types of reader collisions can occur~\cite{wan2021utilizing,golsorkhtabaramiri2022distributed}:
\begin{itemize}
    \item Reader-to-tag collision: This happens when two or more readers try to read a specific tag simultaneously, due to an overlap in their read ranges. As shown in  Figure~\ref{fig:reader-collision}(a), when both Reader1 and Reader2 try to read Tag1 at the same time, a collision occurs since Tag1 falls within the reading range of both readers~\cite{cheng2023enhanced}.
    \item Reader-to-reader collision: This type of collision occurs when one reader's signal interferes with the reception system of other readers. In Figure~\ref{fig:reader-collision}(b), the interference range of Reader1 affects R2's read range, resulting in Reader2's inability to read Tag1~\cite{kumar2025passive,qin2021collision}.
\end{itemize}

As mentioned in paper~\cite{rezaie2023shared}, the problem of re-reading tags is another significant problem in RFID networks. It increases the energy consumption of readers, unnecessarily fills the central server, and reduces throughput. This issue causes much more problems in DRE networks and mobile reader networks. As explained before, in RFID networks, there are several critical issues, foremost among them being the crisis of reader collisions. Additionally, the challenge of duplicate tag readings by readers and the consequential energy consumption quandary further.

This paper is focused on the reader-to-reader collision, reread tags problem, and saving energy. We introduce a centralized method that uses multi-channels to enable simultaneous communications between readers and tags. Each reader adopts the SIFT probability distribution function to select a time slot number~\cite{li2022independent}. The proposed approach defines RFID network efficiency by combining Time Division Multiple Access(TDMA) and Frequency Division Multiple Access(FDMA) mechanisms.

This study increases throughput, decreases energy consumption, and reduces the average waiting time. For doing these tasks, techniques like the SIFT function, exact distance calculation between readers, using the information-sharing phase to reduce rereading tags' information, and leaving the channel on time are applied. Indeed, by preventing tag re-reads and ensuring the release of readers from the communication channel on time, we effectively eliminate reader-to-reader collisions.

\subsection{Contributions}
\begin{itemize}
    \item We propose a method that avoids re-reading tag information with the Information Sharing Phase (ISP). By checking ISP, the readers realize whether tags have already been read or not.
    \item Our proposed method reduces energy consumption compared with benchmark methods. In this method, each reader reduces energy consumption by preventing reading duplicate tags and leaving the communication channel. Our method reduces energy consumption by 52\% compared with benchmark methods. 
    \item We were able to decrease reader collisions and enhance throughput by 26\% compared with benchmark methods. In our method, each reader after reading the tags, shares the tag ID with other readers which causes the throughput to rise. 
    \item Our proposed method reduces the average waiting time with the ISP. In our method, readers at the end of each round can have a lot of tags ID and spend less the average waiting time for reading. The proposed method decreases the average waiting time by 74\% compared with benchmark methods.  
\end{itemize}

\subsection{Organization}
The structure of this paper is as follows: Section \ref{relatedWork} provides a comprehensive review of related works in the field. In Section \ref{Proposed Method}, we present the details of our new proposed method. In Section \ref{Simulation and Evaluation}, we evaluate the proposed method’s performance across various scenarios. Finally, Section \ref{Conclusion and Future Work} presents our conclusions and outlines potential directions for future work.

\section{Related Work}\label{relatedWork}

Industrialization, combined with the integration of emerging technologies, has opened up chances for interaction among devices used in industries, allowing the exchange of information within their working surroundings. This interaction is intended to empower various devices, even if they are different, to pursue shared goals. In RFID systems, readers use different approaches to reduce collisions and communicate effectively with tags. These approaches can be broadly grouped as TDMA, FDMA, and Carrier Sense Multiple Access (CSMA)\cite{golsorkhtabaramiri2022distributed}. Collision problems have led to various suggested solutions, which can be divided into two main methods: distributed and centralized approaches. In the distributed method, readers function independently, without relying on a central server for resource allocation. Conversely, in the centralized technique, a central server oversees reader activities through a wired or wireless network, efficiently supervising resource sharing across the network. This section provides a comprehensive overview of the most important protocols based on these two methods~\cite{rezaie2023radio}.

\subsection{Distributed Protocol}

The Distributed Colour Selection Protocol (DCS) works based on TDMA and distributed methods, dividing time into separate periods with equal time slots, from which readers randomly choose. If there are collisions, readers have to pick an alternative slot randomly and share this update with nearby readers through a kick message. If a reader's chosen time interval overlaps with that of a neighboring reader, the latter must quickly pick a different time slot. Importantly, successful tag readings keep the same time interval in the next period~\cite{waldrop2003colorwave}. On the contrary, the DCS protocol leads readers to select a new time interval after a collision, possibly causing more collisions. To tackle this issue, the Probabilistic Distributed Colour Selection Protocol (PDCS) introduces a probabilistic approach, allowing readers to change their time intervals based on the probability parameter "p". Interestingly, an optimal collision probability of 70\% has been identified for this parameter. Similar to DCS, PDCS is based on the distributed protocol and TDMA, and the use of multiple channels for reader-to-tag communication further improves its efficiency~\cite{gandino2020probabilistic}.

The Colorwave system, which is built upon the DCS protocol, uses a special method where every period consists of different time slots, decided individually by each reader using two sets of thresholds. To notify changes in the number of time slots, readers send kick messages. When these messages are received, nearby readers evaluate their success rates and adjust their time slots based on the threshold pairs. Keeping readers in sync is crucial in this method, and determining the minimum color may result in a high number of connections. While this protocol provides more flexibility in setup compared to DCS, it has a higher overhead and needs an extra signal for color changes. However, its effectiveness is somewhat lower in the initial transmission stages due to the variable time slots. Additionally, it becomes necessary to manage impact messages since collisions may occur when readers have different numbers of time slots~\cite{safa2015distributed}.

The Anti-Collision Protocol for RFID (APR) introduces a spread-out approach based on the Carrier Sense Multiple Access (CSMA) technique. APR enhances communication efficiency by using multiple channels for reader-to-tag interactions, distinguishing between data and control channels to avoid collisions between readers. For communication from reader to tag, the reader that successfully reads the tag’s Identification (ID) sends a beacon message with the tag ID to the control channel. Afterwards, readers that receive this beacon message measure their distance from the reader in the channel by gauging signal transmission power. If considered too close, these readers are deactivated after the current period, ensuring the effective prevention of collisions between readers and tags~\cite{golsorkhtabaramiri2015reader}.

The authors in ~\cite{mafamane2022dmlar} introduced a new method that addresses the challenge of interference in busy RFID networks, which are crucial for the Internet of Things. The proposed solution is a clever system named DMLAR. It employs a unique learning method to predict and avoid collisions between RFID tags and readers. This system is crafted to function effectively even when tags and readers are on the move. The assessment of this system focuses on situations in healthcare where handling important data is crucial. Simulations demonstrate that DMLAR is robust and efficient, especially when dealing with varying numbers of readers and managing resources effectively. This implies that it can assist RFID systems in operating seamlessly in dynamic environments without extensive communication between readers. The approach appears promising for enhancing the performance of RFID systems in places where data is vital and needs to be managed efficiently.

The Distributed Mobile Reader Collision Avoidance protocol (DMRCP) introduces a distributed protocol for RFID networks, using the CSMA mechanism to address issues in environments with mobile readers and distributed tags. It consists of four phases: Waiting for state, Contention, Random delay, and Sharing, to avoid collisions. By employing data and control channels, the protocol optimizes communication intervals and encourages fairness among readers. Sharing tag information within a specified distance range improves the number of readers successfully acquiring tag data, contributing to increased network efficiency and throughput. Experimental validations support the effectiveness of sharing tag data in expanding the reading range between readers~\cite{golsorkhtabaramiri2022distributed}.

The authors proposed MCSMARA, a decentralized MAC protocol for RFID systems that uses a Markov Decision Process (MDP) to optimize reader behavior in dense deployments~\cite{amoah2025mcsmara}. Unlike traditional CSMA or TDMA approaches, MCSMARA enables each reader to dynamically adjust its frequency, time slot, and backoff duration based on utility functions that account for throughput, collision probability, interference, and energy efficiency. The key advantages of MCSMARA include its adaptability to real-time network conditions, improved scalability in dense reader environments, significant collision reduction (up to 30\%), and enhanced throughput (up to 25\%). Additionally, its decentralized design avoids the complexity and overhead of centralized coordination. However, MCSMARA also has limitations: it requires frequent neighborhood information exchange, which may increase control message overhead, and its performance is sensitive to parameter tuning for utility functions and MDP settings. Despite these challenges, MCSMARA offers a robust and scalable alternative to traditional MAC protocols in dynamic and interference-prone RFID scenarios.

\subsection{Centralized Protocol}
The Neighborhood Friendly Reader Anti-Collision Protocol (NFRA) is a centralized TDMA-based system designed for crowded reader environments. Readers are linked to a central server that divides time into slots and informs them through a sequenced packet (AC). After picking a slot randomly from the AC packet, readers get an ordering command (OC) to match their slot number. They send a beacon message to nearby readers and, if collision-free, send an overriding frame packet (OF) to initiate tag reading. This method effectively avoids collisions between readers, ensuring smooth communication~\cite{eom2009efficient}.

The Geometric Distribution Reader Anti-Collision (GDRA) is a TDMA-based system with a centralized mechanism that lessens collisions among readers. By using a geometric probability distribution function, readers randomly pick time slots, decreasing the chances of collisions. The winner gains access to the tag and communicates data in turn to the central server through the low-level reader protocol (LLRP). GDRA ensures effective tag reading and control over interference, promoting smooth communication~\cite{bueno2012geometric}.

The NFRA Contention (NFRA‐C) protocol improves upon the initial NFRA by introducing a counter-based strategy. Every reader keeps a counter to monitor successful tag communication, incremented with each connection. Counters are shared through beacon messages. In the event of a collision, readers compare their counters, and the one with the lower value gets remote access. This prioritization guarantees higher throughput and fairness in crowded RFID networks. NFRA‐C performs better than NFRA by 15\%, ensuring enhanced efficiency, particularly in situations with numerous neighbors and collision possibilities~\cite{nawaz2015nfra}.

The Distance Reader Collision Avoidance (DRCA) protocol, as explained in~\cite{golsorkhtabaramiri2017distance}, acts as a centralized anti-collision system designed for busy fixed or mobile reader environments. Its main goal is to allow multiple readers to operate simultaneously without experiencing collisions. Based on TDMA, the protocol uses a SIFT distribution to independently determine intervals. When checking channel occupancy, a reader adjusts its interval if it's beyond the active reader's range, then switches to a randomly chosen channel. The protocol ensures collision avoidance by sending beacon messages on new channels and refrains from operating if interference with active readers is detected. This strategic approach reduces collisions, leading to increased throughput in RFID networks.

The Fairness Reader Collision Avoidance 1 (FRCA1) approach, similar to NFRA, establishes coordination between the server and readers. By combining FDMA and TDMA mechanisms, this approach achieves improved throughput and reduced network waiting time, resulting in better fairness. Operating at different frequencies simultaneously prevents signal interference between readers, leading to fewer collisions among readers and more successful readings, ultimately enhancing overall throughput. The Fairness Reader Collision Avoidance 2 (FRCA2) method partly addresses the reader-to-tag collision issue found in FRCA1. However, to reduce reader-tag collisions, certain readers are restricted from reading, potentially causing a decrease in throughput compared to FRCA1~\cite{rezaie2018fair}.

The NFRA-adaptive interrogation capacity protocol emerges as an extension of the foundational NFRA framework. This innovative approach empowers readers to adjust their interrogation duration based on the number of tags within their respective interrogation zones. The introduction of sub-rounds allows readers to conclude interrogation conditions after completing tag interrogations, prompting notifications to neighboring readers. Following this, surrounding readers can decide whether to re-enter the competition, guided by the status of their nearby counterparts. The strategic allocation of a time interval between the AC signal and the first OC signal enables readers still involved in tag interrogations to continue without facing additional competition, as highlighted in reference~\cite{li2019nfra}.

This study~\cite{rezaie2023radio} introduces an inventive RFID system approach that merges time division and the simultaneous use of multiple frequencies, incorporating TDMA and FDMA protocols. Guided by a central server, readers prevent collisions by selecting intervals through the sift distribution function. This strategy enhances the network's operational capacity, avoids collisions, and conserves energy during information-sharing phases. In tackling the challenges of reader collisions in Industry 4.0, the IRAP method utilizes TDMA and FDMA techniques while adhering to RFID standards. The sharing of information among readers boosts productivity and resource savings, with simulation results showcasing the method's efficiency in terms of throughput compared to existing approaches. The research represents a significant contribution to optimizing RFID systems in high-density environments.

The authors proposed RDDS, an improved anti-collision algorithm designed to address the limitations of traditional ALOHA-based protocols under the capture effect~\cite{le2025improved}. As a centralized protocol, RDDS reduces the reader-to-tag communication overhead by simplifying the command set and modifying the tag-side state machine. The main contribution lies in its ability to re-identify weak signal tags more efficiently, even when the capture effect occurs. RDDS outperforms its predecessor DDS in terms of fairness and weak-tag recognition without significantly sacrificing overall throughput. However, it omits certain optimization steps like the Divide command, which may lead to slightly lower performance when capture effects are not present. Overall, RDDS provides a more robust and bandwidth-efficient solution for small-scale or capture-prone RFID deployments.

\section{\textbf{Proposed Method} }\label{Proposed Method}
The proposed method in this paper is the Intelligence and Efficient Reader Anti-collision Protocol (IE-RAP), which is a centralized protocol to address communication between the server and readers. At the beginning of each round, the server broadcasts message 'A' to all readers. This message includes an active range for time slots (from 1 to the maximum number (MN)) and another range for the channel number (from 1 to F). Readers then independently generate random values for the channel number and time slot number using the random and SIFT distribution functions, respectively. Subsequently, the server sends message 'C' to all readers, containing the current time slot number. Readers compare their time slot numbers ($k$) with the value in message 'C'; if they match, they start to broadcast beacons to detect collisions. When multiple readers share the same time slot and channel number, a beacon collision occurs. Then several situations can occur: 
\begin{itemize}[leftmargin=*]
    \item If there were only two readers, the readers calculate the distance ($D$) between each other. Then one of the following situations can happen:
    \begin{itemize}[leftmargin=*]
        \item If $D$ was less than \textit{two} times the \textit{read range}, the reader with a lower number of successful readings ($S$) takes over the channel. After that, it checks whether it has already received tags ID from other readers or not. If tags have not been read, the reader starts reading them. Then at the end of the round, there is the information-sharing phase (ISP), where the reader shares the tag ID and its $S$ (that one should be added) in the ISP part.     
        However, if the reader already has tags ID, it listens to the channel, awaiting the broadcast of the 'C' message from the server. After hearing it leaves the channel so that other readers can use that channel in the next time slot and the reader goes to sleep to save energy. At the end of the round, the server broadcasts an 'SH' message, so that all readers wake up and receive information from the ISP part. After sharing this information, every reader should update their memory.
        
        \item If the D between \textit{two} readers is more than twice the \textit{read range}, the reader with a lower number of successful readings ($S$) takes over the channel. After that, it checks whether it has already received tags ID from other readers or not. If tags have not been read, the reader starts reading them. Then, at the end of the round, there is the information-sharing phase (ISP), where the reader shares the tag ID and its $S$ (that one should be added) in the ISP part.     
        However, if the reader already has tags ID, it listens to the channel, awaiting the broadcast of the 'C' message from the server. After hearing it leaves the channel so that other readers can use that channel in the next time slot and the reader goes to sleep to save energy. At the end of the round, the server broadcasts an 'SH' message, so that all readers wake up and receive information from the ISP part. After sharing this information, every reader should update their memory.
        
The reader who has the higher $S$, adds one to its slot number and waits to hear a new 'C' message from the server until it participates in the competition again. 
    \end{itemize}
\end{itemize}
\begin{itemize}[leftmargin=*]
    \item If the number of readers is more than \textit{two}, everyone will go to sleep and wait until the start of the new round and receive the 'A' message from the server.
\end{itemize}

This paper proposes an innovative method for improving the efficiency and performance of RFID networks with shared channels. The proposed method involves readers sharing information about the number of successful readings (S) and the tags' ID that they have read at the end of each round. By doing so, several benefits are realized. Firstly, readers avoid reading duplicate tags, leading to energy savings. Secondly, when they identify an already-read tag, they immediately go to sleep during the next time slot so that the rest of the readers can use the channel. This procedure increases the likelihood of other readers successfully reading tags in the same round, reducing energy consumption, and enhancing network throughput. Moreover, the shared channel allows each reader to obtain tag IDs much faster, resulting in a significant reduction in the average waiting time for reading tags. These improvements contribute to a more efficient and resource-saving RFID network, making it a promising solution for various practical applications. 

\subsection{Definitions}
\noindent In the proposed method, we will use the following definitions:
\begin{enumerate}[leftmargin=*]
    \item\textbf{Distance Between Readers:} For calculating distance between readers, we borrow Eq.~\eqref{Distance-formula-1} from the method proposed in ~\cite{zhang2013simulation}. This approach enables readers to estimate the distance between themselves and other readers using signal interference analysis rather than requiring direct knowledge of all parameters from other readers. This approach is based on RFID interference models like the unit disk graph model and the additive interference model. In this equation, $D$ is the distance between the \textit{two} readers (R1 and R2), which is shown in Figure~\ref{fig:readerdistance}.
    
%%************************************************************    
\begin{equation}\small\label{Distance-formula-1}
     D^{\propto}=\frac{(P_R)*(G_{R1})*(G_{R2})}{(K_0)*(I_R)}  
\end{equation}
%%************************************************************
 Where, $\alpha$ is the path loss exponent; $P_R$ is the reader transmission power; $G_R$ is the reader antenna gain, and  $K_0$ is the coefficient of channel path loss and power ratio in the bandwidth, while $I_R$ denotes the total interference that the reader receives. Readers operate under homogeneous network assumptions, meaning all readers have predefined and equal transmission power and antenna gain. Since these parameters are typically standardized across RFID systems, a reader can assume that other readers share the same $P_R$ and $G_R$ values. Readers do not explicitly communicate their position but infer distance based on received signal strength (RSSI) ($I_R$) and path loss models ($K_0$).

%%************************************************************ 
\begin{figure}[!htbp]
\centering 
\includegraphics[width=0.95\columnwidth]{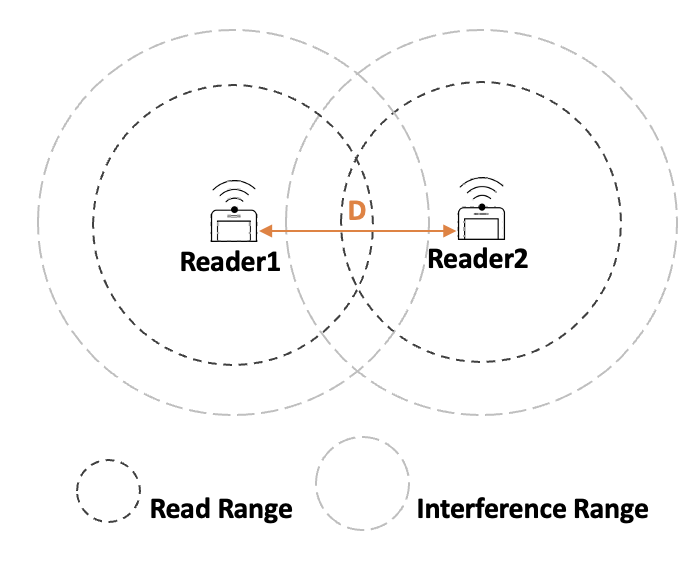}
\caption{ \small Distance Between Two Readers }
\label{fig:readerdistance}
\end{figure}
    \item \textbf{SIFT Distribution Function:}
    In the NFRA protocol (as we reviewed in the related work), readers randomly select one of the time slots based on the uniform distribution function. Based on this function, the probability of a collision in each time slot is the same for rival readers. Rival readers are those which we have in the reading range and interference range of each other and operate in one channel~\cite{bueno2012geometric}. 
In the $CSMA⁄P^*$ protocol~\cite{tay2004collision}, network nodes use the non-uniform $P^*$ probability distribution according to Eq.~\eqref{probability-distribution} to select competing slots. This function minimizes collisions between competing readers and maximizes the probability of selecting a time slot by just one reader.
\begin{equation}\small\label{probability-distribution}
     P^*_k=(\frac{1-f_{K-k(R)}}{R-f_{K-k(R)}}) \left(  1 - P^*_1 - P^*_2 -\ldots \quad - P^*_K{-1}\right)
\end{equation}
Given Eq.~\eqref{recursive-function}, $f_{K-k}(R)$ is a recursive function for 1$<$=k$<$=K : 
\begin{equation}\small\label{recursive-function}
    f_{K-k}(R) = (\frac{R-1}{R-f{K-k-1}(R)})^{R-1}
\end{equation}
Where 2$<$=k$<$=K , R$>$=2 and $f_1(R)=0$ . 
For the $CSMA⁄P^*$  protocol to be used in a dense environment of the RFID system, each reader must be able to estimate the number of neighbors. But if the reader does not know the number of its neighbors, the SIFT distribution probability ($P_k$ ) function is used to select competitive intervals according to Eq.~\eqref{SIFT-distribution-probability}~\cite{global2008epc}.
\begin{equation}\small\label{SIFT-distribution-probability}
     P_k =\frac{(1-\alpha)\alpha^K}{1-\alpha^K} \alpha^{-k} 
\end{equation}
Eq.~\eqref{uniform-probability-distribution} holds for 1$<$=k$<$=K, 0$<$ $\alpha$ $<$1 and $\alpha$ = $M^{\frac{-1}{K-1}}$; where $M$ is the maximum number of competing readings~\cite{global2008epc}. When $\alpha$ = 1 and $M=1$, the Eq.~\eqref{uniform-probability-distribution} corresponds to a uniform probability distribution function:
\begin{equation}\small\label{uniform-probability-distribution}
     \lim_{\alpha \to 1 } P_k = \frac{1}{K}
\end{equation}
In the SIFT probability distribution function, the probability of selecting higher time slots will increase. Consequently, this approach leads to an increased probability of individual readers selecting lower time slots. In the SIFT probability distribution function, the probability of winning the competition in the presence of $R$ neighbors for a reader is computed by Eq.~\eqref{SIFT}~\cite{global2008epc}: 
\begin{equation}\small\label{SIFT}
     P_c(R) = R \sum_{k=1}^{K-1} P_k ( 1 - \sum_{z=1}^{K} Pz )^{R-1}
\end{equation}

\end{enumerate}
%%************************************************************

\subsection{Protocol Algorithm}

The pseudocode of the proposed method is presented in  Alg.~\ref{alg:newmethod}. In lines 1-5, the server broadcasts message 'A' and the readers select the frequency number and time slot based on the Random and SIFT functions. Lines 6 and 7 show that if the reader’s time slot number (K) is equal to the number mentioned in message 'C' (P), the reader will send a beacon message.

Lines 8-30 show that if there were two conflicting readers and the distance between them was less than twice the read range, that means they are neighbors. In lines 10-24, we defined a $HandleTagReading$ procedure that the reader whose number of successful readings ($S$) is less than the rival reader's $S$ succeeds in getting the channel. If the tag has not been read before, the reader reads the tag and puts the tag's ID in the Information-sharing phase (ISP). However, if the tag has already been read, the reader leaves the channel in the next time slot. So that other readers can use that frequency in the next slots and wait for the 'SH' message from the server to read the information in the ISP. In line 25, if distance between two reader was less than reader's read range, call the $HandleTagReading$ procedure. Else, in lines 28-30, a reader with a lower number of successful readings ($S$) waits until the server broadcasts message 'A'.

%%************************************************************

\begin{algorithm}[H]
\caption{\small \textbf{IE-RAP Algorithm }}
	\footnotesize
\label{alg:newmethod}

\begin{algorithmic}[1]
\color{black}
\footnotesize 

\If{(Receive 'A' message from the server)} %1 
\State{$ F \leftarrow \text{Random among [1, F] as a channel number}$}  %2
\State{$ K \leftarrow \text{SIFT function among [1, MN] as a time slot number} $} %3
\State{$\text{Wait and Received 'C' message from the server}$}%4
\State{$P \leftarrow \text{Extract the number in 'C' message}$} %5
\If{(K == P)} %6
\State{$\text{Broadcasts a beacon in the channel}$} %7
\If{ A collision of the beacons between 2 readers is detected} %8
\State{$ D \leftarrow \text{Readers calculate their distance}$} %9
\Procedure{HandleTagReading}{S(reader), S(rival reader)} % Function to handle tag reading
    \If{ ($S$(reader) $<$ $S$ (rival reader))} %1
        \If{ The reader reads tags that were not read before} %2
            \State{$\text{The reader starts reading tags}$} %3
            \State{$ S \leftarrow \text{$S$ + 1 }$} %4
            \State{$\text{Shared the tag information and $S$ in the ISP}$} %5
        \Else  %6
            \State{$\text{The reader avoids re-reading}$} %7 
            \State{$\text{Waits for the 'C' message}$} %8 
            \State{$\text{Leaves the channel}$} %9
            \State{$\text{Waits for the 'SH' message}$} %10
            \State{$\text{Reads the ISP}$} %11
        \EndIf  %12
    \EndIf %15
\EndProcedure
\If{ D $<$= (2 $\times$ Reader’s read range)} %10
\If{\State{\Call{HandleTagReading}{S(reader), S(rival reader)}}}
\Else  \Comment{S(reader) $>$ S (rival reader))}%23
\State{$\text{ Wait for the next 'A' message}$} %24
\EndIf %25
\Else \Comment{ $D$ $>$ (2 $\times$ Reader’s reading range}) %26
\If{\State{\Call{HandleTagReading}{S(reader), S(rival reader)}}} %26
\Else  \Comment{S(reader) $>$ S (rival reader))} %39
\State{$ K \leftarrow \text{SIFT function among [1, MN] as a time slot number} $} %40
\State{$K\leftarrow \text{$K$ + 1 } $} %41
\State{$\text{Waits for the next 'C' message from the server }$} %42
\EndIf %43
\EndIf %44
\Else \Comment{ More than 2 readers can read} %45
\State{$\text{Waits for the next 'A' message from the server}$} %46
\EndIf %47
\Else \Comment{ ($K$ != the number in 'C' message)} %48
\State{$\text{Waits for the next 'C' message from the server}$} %49
\EndIf %50
\State{$\text{The server broadcasts the 'SH' message to wake readers to share }$} %51
\State{$\text{Receive the new tag's ID in the ISP}$} %52
\EndIf %53

\end{algorithmic}
\end{algorithm}

%%************************************************************

In lines 31-39, if the distance between readers is more than twice the read range, call the $HandleTagReading$ procedure that the reader that has less number of successful readings ($S$) succeeds in getting the channel. If the tag has not been read before, the reader reads the tag and shares the tag's ID in the information-sharing phase (ISP). However, if the tag has already been read, the reader leaves the channel in the next time slot. So that other readers can use that frequency in the next slots. The reader with a lower number of successful readings ($S$) in the same round selects a new frequency, increases its previous slot number by one unit, and waits to receive the 'C' message.

From lines 40-42, if the number of conflicting readers is more than two, they leave the channel until the message 'A' is broadcast again. From lines 43-48, if the reader's $K$ number is not the same as the slot number, wait for the next slot number. In line 46, the server broadcasts the 'SH' message to wake readers to share and receive the new tag's ID in the ISP.\\

\subsection{Toy example}
\begin{figure}[!htbp]
\centering 
\includegraphics[width=0.9\columnwidth]{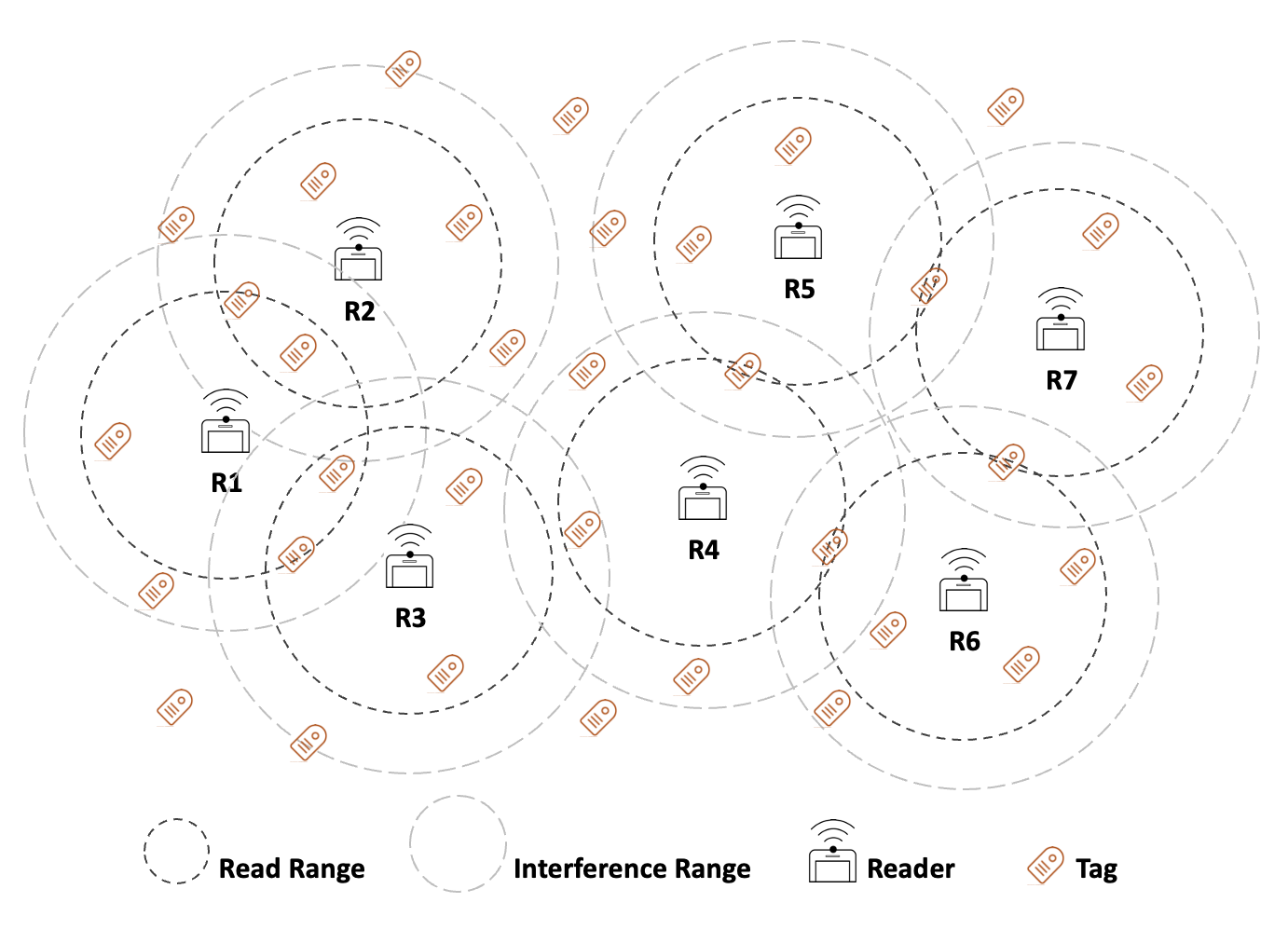}
\caption{ \small Example of Network }
\label{fig:Example of Netwrok}
\end{figure}

We apply the proposed method for the RFID network shown in Figure~\ref{fig:Example of Netwrok}. As we can see, there is overlap between the following read ranges: 

%%************************************************************
\begin{table}[!htbp]
\begin{center}
%\footnotesize
%\begin{adjustbox}{max width=\textwidth}
\begin{tabular}{cccc}
 R1 and R2, & R1 and R3,& R3 and R4, & R4 and R5 \\
R4 and R6, & R7 and R5,& R7 and R6 &\\
\end{tabular}\vspace{-25px}
%\end{adjustbox}
\end{center}
\end{table}
%%************************************************************

\begin{figure}[!htbp]
\centering 
\includegraphics[width=0.95\columnwidth]{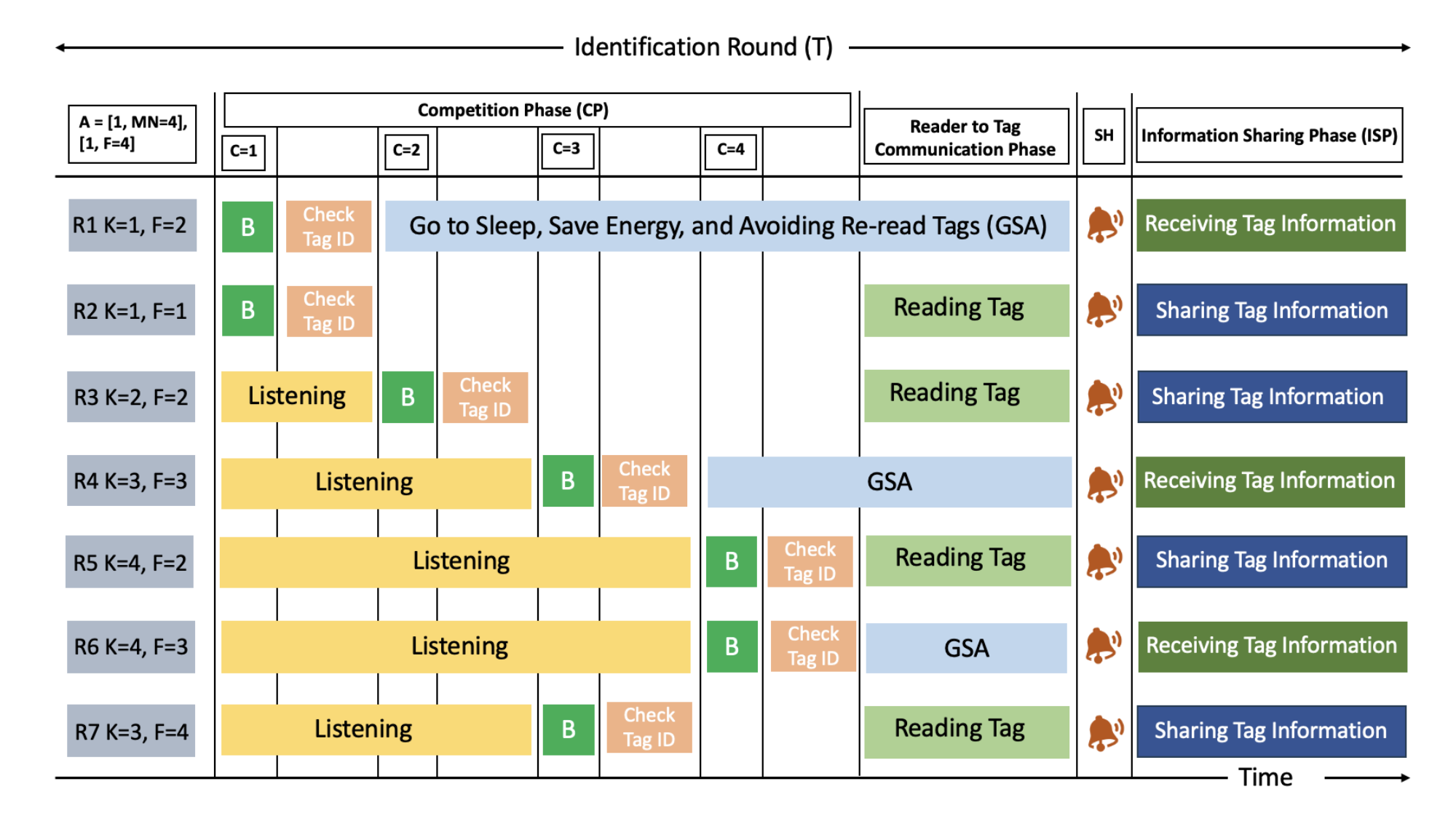}
\caption{ \small Example of New Method Procedure }
\label{fig:New Method}
\end{figure}

\noindent Figure~\ref{fig:New Method} shows one round of applying the proposed method on the RFID network which is presented in Figure~\ref{fig:Example of Netwrok}. At the initial stage of each round, the server broadcasts a message 'A', and then all readers choose a channel number (F) between [1, 4](suppose we have 4 channels.) and a time slot number (K) between [1, MN=4].  Readers listen to the channel when the 'C' message heard matches their timeslot number (K) and send a beacon message (B) to see if a collision occurs.

R1 and R2 chose the same time slot number (K=1), then they sent beacons. Because they are not on the same channel (different F values), they do not disturb each other, and they can use their channels. Each of these readers checks before starting to read the tag to see if the tag has already been read by other readers or not. If the tag has already been read, the readers have the tag information in the information sharing phase (ISP) and will not read the tag again. This issue is very important because the readers are mobile and their neighborhoods have changed. We suppose R1 realizes that it has the information of the tags around it and waits for the next 'C' message. After getting the 'C' message, it goes to sleep. So, R1 does not waste energy reading duplicate tags and frees up channel number 2 for the next time slots. In this way, more readers can succeed in reading. However, R2 does not recognize the tag to start reading. Even though R3 overlaps with R1, they have chosen a common channel (F=2). Because R1 has left the channel in the previous time slot, R3 sends a beacon to find the empty channel and start reading. 

R4 selects time slot 3 and, after sending the beacon, finds the empty channel. However, it has the tag information and leaves the channel in the next time slot so that other readers can use the channel. R7 has selected time slot 3 and finds the empty channel, and because it does not have the tags information, it starts reading tags. R5 and R6 have chosen time slot 4, they both send beacons and can read tags at different channels (R5 in F=3 and R6 in F =2). R5 also sends beacons, and since R4 has left channel 3 in time slot 3, it finds the empty channel. It does not have the tag information, so it starts reading the tags. R6, after checking the tag information, finds out that it has tag information and leaves the channel. All readers wake up after hearing the 'SH' message from the server, share their information, and receive other's information. All readers wait for the start of the next round by the server.

As can be seen, mobile readers engage with a variety of tags at different times. Concurrently, during the Information Sharing Phase (ISP), they can effortlessly read and access a wealth of tag information. This method not only conserves energy but also contributes to an abundance of information acquisition. Consequently, network throughput increases, and the waiting time on the network decreases. Moreover, the duplication of tag readings is minimized.

\section{\textbf{Simulation and Evaluation} }\label{Simulation and Evaluation}
This section presents the simulation and evaluation results of the proposed method (IE-RAP) using the R2RIS software~\cite{ferrero2013simulating}. R2RIS is a specialized simulator designed to assess the efficiency and impact of reader collisions in RFID systems. The RFID system under consideration has readers with Bistatic antennas, distributed randomly within a square environment spanning 1000 square meters. Each reader operates at an output power of 2.3 watts EIRP. The readers' read range is up to 10 meters and their interference range is up to 1000 meters~\cite{jiang2012efficient}. Simulation parameters of the existing protocols and IE-RAP are in Table 1, while other necessary parameters are taken from this paper~\cite{eom2009efficient}.\\

\begin{table}[h]
\centering
\caption{Simulation parameters}
\label{tab: Simulation Parameters}
\begin{tabular}{c|c|c}
\hline
Methods & Parameters & Values \\
\hline
\multirow{3}{4em}{IE-RAP} & Reader Reading & 0.46 Seconds \\ 
& Beacon packet & 0.3 Millisecond \\ 
& A packet & 2.83 Period time \\ 
& C Packet & 2 Millisecond \\ 
& SH Packet & 1 Millisecond \\
& T (slot) & 5 Millisecond \\
& Number of time period & 128 \\ 
\hline
\multirow{3}{4em}{GDRA~\cite{bueno2012geometric}} & Reader Reading & 0.46 Seconds \\ 
& Beacon packet & 0.3 Millisecond \\ 
& AC packet & 2.83 Period time \\ 
& OC Packet & 1 Millisecond \\ 
& T(slot) & 5 Millisecond \\ 
\hline
\multirow{3}{4em}{NFRA~\cite{eom2009efficient}} & Reader Reading & 0.46 Seconds \\ 
& Beacon packet & 0.3 Millisecond \\ 
& AC packet & 2.83 Period time \\ 
& OC Packet & 1 Millisecond \\ 
& OF Packet & 0.3 Millisecond \\  
\hline
\multirow{3}{4em}{FRCA~\cite{rezaie2018fair}} & Reader Reading & 0.46 Seconds \\ 
& Beacon packet & 0.3 Millisecond \\ 
& SO packet & 2.83 Period time \\ 
& EO Packet & 1 Millisecond \\  
\hline
\multirow{3}{4em}{DMRCP~\cite{golsorkhtabaramiri2022distributed}} & Reader Reading & 0.46 Seconds \\ 
& Beacon packet & 5 Millisecond \\ 
& Contention Windows (CW) & 5\\
\hline
\end{tabular}
\end{table}
\vspace{-5px}
For comparative analysis, the protocols are assessed based on three key parameters: throughput, the average waiting time of each reader, and energy consumption. Throughput refers to the number of tags read per unit of time, and the additional time required for extra actions in IE-RAP was accounted for in throughput calculations. Meanwhile, average waiting time represents the average time taken by all readers to read a tag. These metrics allow us to measure the performance and efficiency of the proposed method with existing protocols and the energy consumption as the amount of energy consumed by all readers in the network~\cite{ferrero2013simulating}.

The energy consumption of each reader was considered following paper~\cite{golsorkhtabaramiri2019comparison}, which includes the total energy of successful reading of tags ($E_{Read(i)}$)  plus the total energy consumed to send control packets ($E_{Send(i)}$) and the total energy consumed to receive control packets ($E_{Receive(i)}$), where $i$ is the identifier of each reader. Finally, the total energy consumption of each reader ($E_{Reader(i)}$) was obtained from Eq.~\eqref{E-reader}:

\begin{equation}\small\label{E-reader}
    E_{Reader(i)} = \sum E_Read(i) + \sum E_Send(i) + \sum E_Receive(i)
\end{equation}

On the other hand, $E_{Send(i)}$ is obtained from Eq.~\eqref{E-send}, where $P_{Send(i)}$ is the transmission power and $T_{Send(i)}$ is the time taken to transfer the packet.

\begin{equation}\small\label{E-send}
    E_{Send(i)} = P_{Send(i)} \times T_{Send(i)}
\end{equation}

$E_{Receive(i)}$ is obtained from Eq.~\eqref{E-receive}, where $P_{Receive(i)}$ is the transmission power and $T_{Receive(i)}$ is the time taken to transfer the packet.

\begin{equation}\small\label{E-receive}
    E_{Receive(i)} = P_{Receive(i)} \times T_{Receive(i)}
\end{equation}

$E_{Read(i)}$ is obtained from Eq.~\eqref{E-read}, where $P_{Read(i)}$ is the transmission power and $T_{Read(i)}$ is the time taken to transfer the packet.

\begin{equation}\small\label{E-read}
    E_{Read(i)} = P_{Read(i)} \times T_{Read(i)}
\end{equation}

Finally, the Network Energy Consumption is obtained from Eq.~\eqref{Energy-consumption}, which is the unit of energy J.

\begin{equation}\small\label{Energy-consumption}
    Energy-Consumption = \sum E_{Reader(i)} 
\end{equation}

Several scenarios have been considered to demonstrate the performance of the proposed protocol (IE-RAP), the results of which have been compared with NFRA~\cite{eom2009efficient}, FRCA1~\cite{rezaie2018fair}, FRCA2~\cite{rezaie2018fair}, GDRA~\cite{bueno2012geometric}, and DMRCP~\cite{golsorkhtabaramiri2022distributed} protocols.

\begin{figure}[!htbp]
\centering 
\includegraphics[width=0.95\columnwidth]{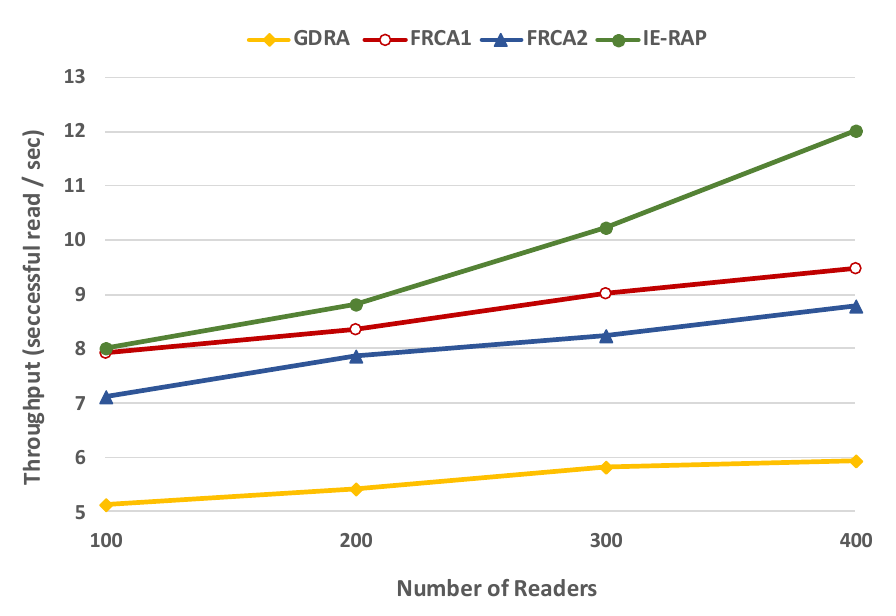}
\caption{ \small Throughput of the evaluate condition in Scenario 1.}
\label{fig:throughput 1}
\end{figure}

In the first scenario, IE-RAP is compared with three protocols, namely FRCA1, FRCA2, and GDRA, in a 4-channel mode. In our experiments, 100, 200, 300, and 400 fixed readers are randomly distributed in a space of 1000 square meters. The number of competitive slots is 128 and the collision rate in this environment was approximately 100\%. As shown in Figure~\ref{fig:throughput 1}, the performance of IE-RAP has been compared with the FRCA1, FRCA2, and GDRA protocols. The throughput of IE-RAP was better than that of other protocols at all evaluation points. This better throughput is because, in IE-RAP, each reader shares the tags' ID with others after reading the tags. So, the readers read more tags in each round, and the throughput rises when the number of readers increases.

\begin{figure}[!htbp]
\centering 
\includegraphics[width=0.95\columnwidth]{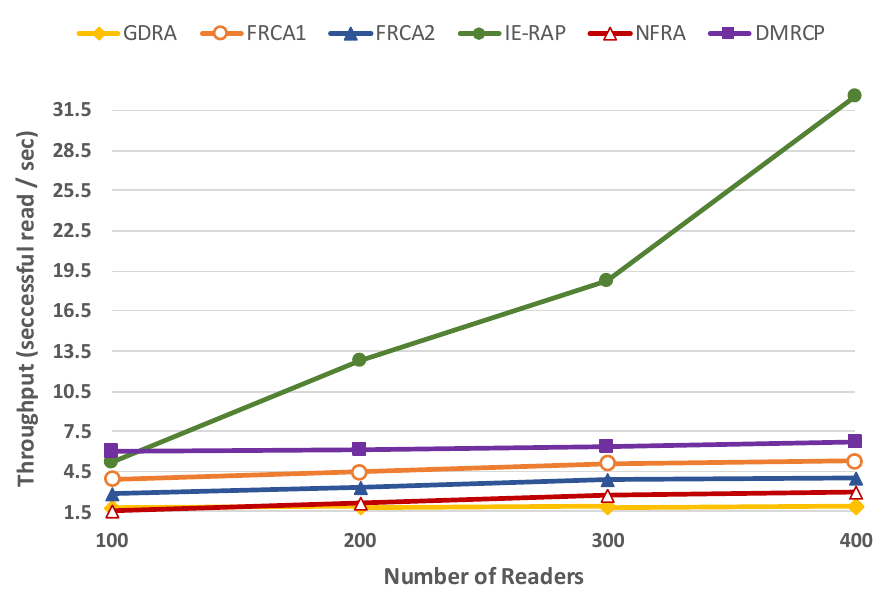}
\caption{ \small Throughput of the evaluate condition in Scenario 2 }
\label{fig:Throughput 2}
\end{figure}

The second scenario compares IE-RAP in single-channel mode with the NFRA, DMRCP, FRCA1, FRCA2, and GDRA protocols. In this scenario, there are 100, 200, 300, and 400 fixed readers which are randomly distributed in a space of 1000 square meters. The number of competitive slots is 128. As shown in Figure~\ref{fig:Throughput 2}, the throughput of IE-RAP has been compared with the NFRA, FRCA1, FRCA2, and GDRA protocols. IE-RAP, even in single-channel mode, has much better results than the NFRA, FRCA1, FRCA1, FRCA2, and GDRA protocols. In single-channel scenarios, the readers have more neighbors, and as a result, the readers are more likely to be close to each other. Because in IE-RAP, the readers can sleep on time and the channel is freed for other readers in the next time slots. The readers can share the tags' ID at the end of the round.

\begin{figure}[!htbp]
\centering 
\includegraphics[width=0.95\columnwidth]{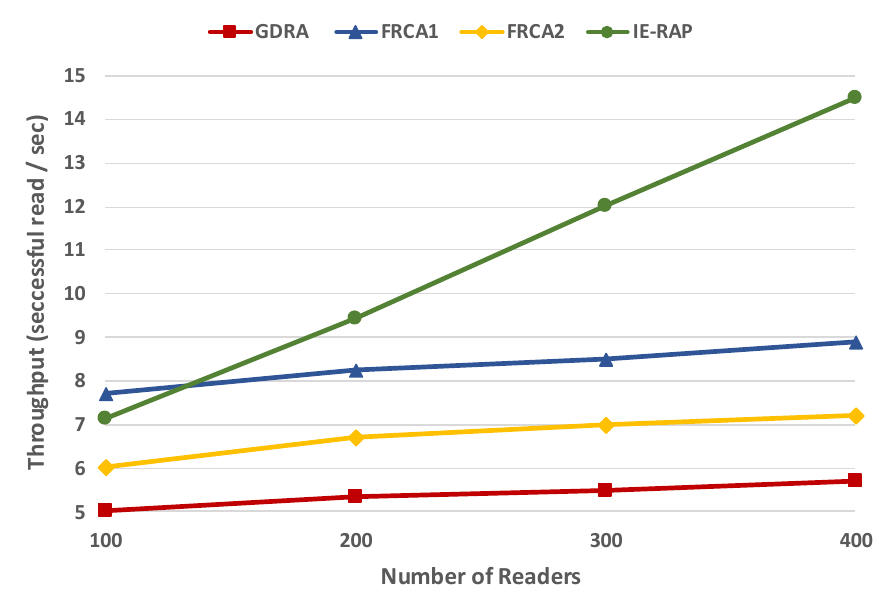}
\caption{ \small Throughput of the evaluate condition in Scenario 3 in 4-channel}
\label{fig:throughput 3}

\end{figure}

\begin{figure}[!htbp]
\centering 
\includegraphics[width=0.95\columnwidth]{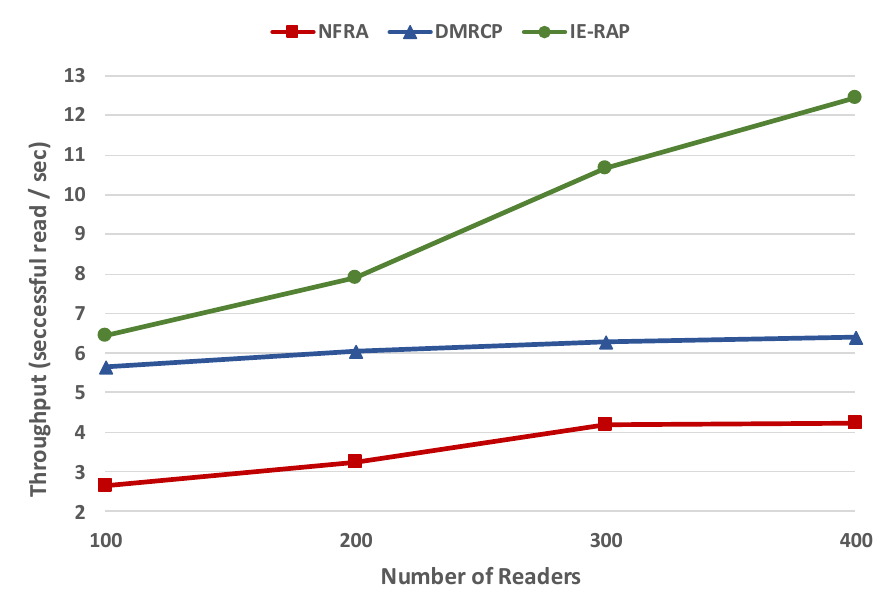}
\caption{ \small Throughput of the evaluate condition in Scenario 3 in single-channel}
\label{fig:throughput 4}
\end{figure}

The third scenario compares IE-RAP with three protocols, namely FRCA1, FRCA2, and GDRA in 4-channel mode, and with two protocols, namely NFRA and DMRCP in single-channel mode. In this scenario, there are 100, 200, 300, and 400 mobile readers which are randomly distributed over an area of 1000 square meters. Also, the number of time slots is 128. Due to the presence of mobile readers, the average number of readers that are very close to each other, in each period, is different. In Figure~\ref{fig:throughput 3}, we present the result of the throughput comparison among IE-RAP, FRCA1, FRCA2, and GDRA in the 4-channel mode and In Figure~\ref{fig:throughput 4}, we represent the result of throughput comparison among  IE-RAP, NFRA, and DMRCP.  In both Figures, IE-RAP’s throughput was better than other protocols in all evaluation points. Because the readers in IE-RAP can share tag ID and leave the channel at the end of each round. It increases other reader's chances of gaining the channel.

\begin{figure}[!htbp]
\centering 
\includegraphics[width=0.95\columnwidth]{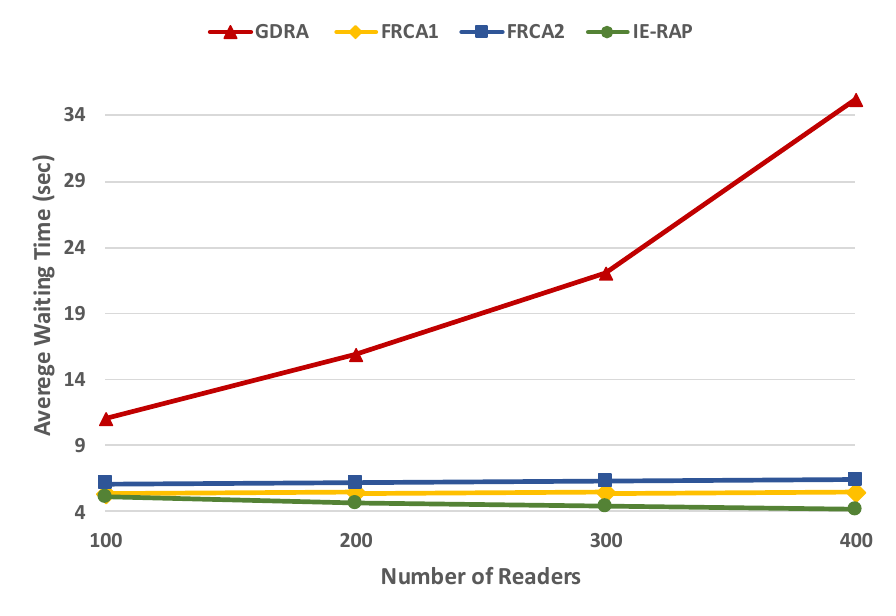}
\caption{ \small Average waiting time of the evaluate condition in Scenario 4 in 4-channel }
\label{fig:average-waiting-time1}
\end{figure}

\begin{figure}[!htbp]
\centering 
\includegraphics[width=0.95\columnwidth]{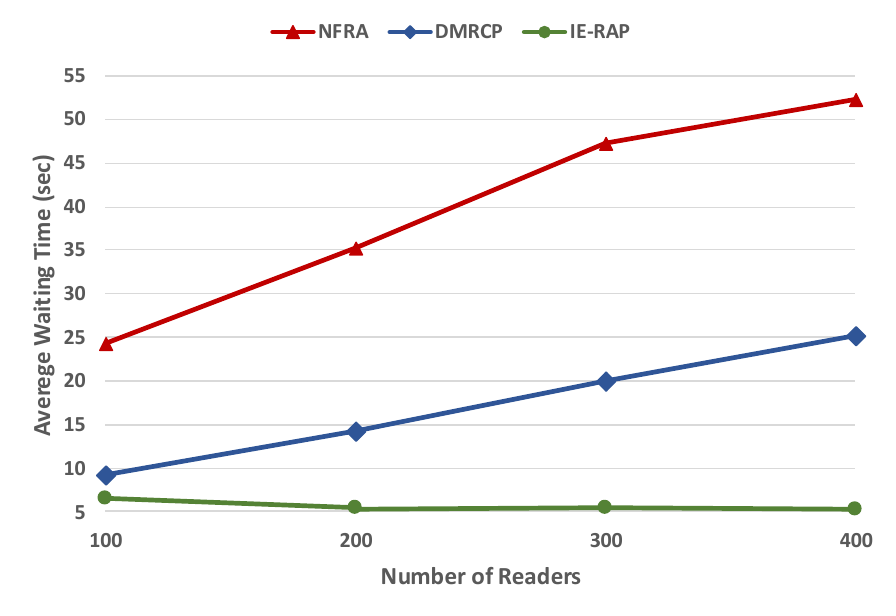}
\caption{ \small Average waiting time of the evaluate condition in Scenario 4 in single-channel }
\label{fig:average-waiting-time2}
\end{figure}

The fourth scenario compares IE-RAP with GDRA, FRCA1, and FRCA2 protocols in 4-channel mode and NFRA and DMRCP in single-channel mode. In this scenario, there are 100, 200, 300, and 400 mobile readers which are randomly distributed over an area of 1000 square meters. The number of competitive slots is 128. The results, provided in Figure~\ref{fig:average-waiting-time1} are among IE-RAP, GDRA, FRCA1, and FRCA2 protocols in 4-channel and in Figure~\ref{fig:average-waiting-time2} are among IE-RAP, NFRA, and DMRCP protocols in single-channel, show that the average waiting time of each reader in IE-RAP was less than other methods. Because in IE-RAP, readers at the end of each round can have many tags' ID and spend less the average waiting time for reading.

\begin{figure}[!htbp]
\centering 
\includegraphics[width=0.95\columnwidth]{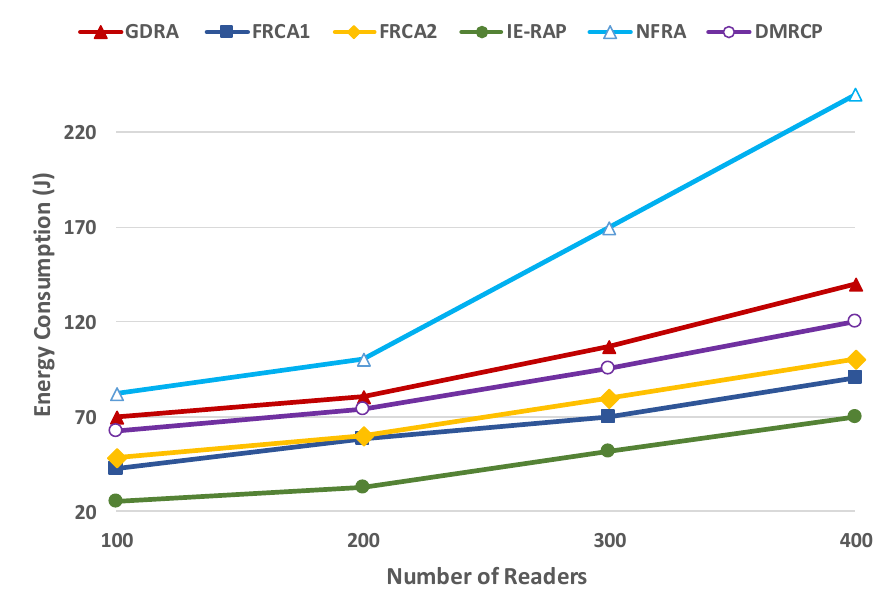}
\caption{ \small Energy Consumption of the evaluate condition in scenario 5}
\label{fig:energy consumption1}
\end{figure}

The fifth scenario compares IE-RAP with GDRA, FRCA1, and FRCA2 protocols in 4-channel mode and NFRA and DMRCP in single-channel mode. In this scenario, there are 100, 200, 300, and 400 mobile readers which are randomly distributed over an area of 1000 square meters. The number of competitive slots is 128. The results, provided in Figure~\ref{fig:energy consumption1}, show that IE-RAP consumes less energy than other simulated methods. At the end of each round, the readers can have more tags' ID, and they can go to sleep as soon as they realize that the tag has already been read. The energy consumption of the readers is reduced, because they do not read the duplicate tags.

\section{\textbf{Conclusion and Future Work} }\label{Conclusion and Future Work}

RFID systems are one of the most important ways to automate mobile networks, as defined by IoT and the use of new technologies. If multiple readers use the channel simultaneously, a collision problem occurs, which is due to the lack of optimal management of channel access control. This is an important difficulty affecting the throughput, efficiency, and energy consumption. One of the most important goals of RFID systems is to increase throughput while avoiding reader collision and decrease energy consumption and average waiting time. The IE-RAP method reduces the collision problem by applying the TDMA and FDMA techniques, preventing the rereading of tags, and sharing tag information. Also, in this method, to increase productivity and resource savings, the readers obtain information about their surrounding readers by sharing information, which increases the throughput when the number of readers increases in the environment. This method, by preventing the rereading tags, decreases energy consumption. The efficiency of the proposed method has been compared and evaluated by simulations and comparisons with previously presented methods, the results of which were significant in terms of throughput, average waiting time, and energy consumption.

For future studies, we intend to allocate frequency channels and time slots to each reader using machine learning techniques more efficiently, reduce the number of possible unemployed time slots, and use successful readings of readers to prioritize them in their potential competition for time slots.

%\section*{Acknowledgment}
\Urlmuskip=0mu plus 1mu\relax
\bibliographystyle{elsarticle-num}
\bibliography{ref}
%\balance
%\clearpage
%\pagebreak

\newpage
\section*{Biographies}\label{sec:11}
\vspace{5px}

\noindent\begin{minipage}{0.13\textwidth}
\includegraphics[width=1.15in,height=1.15in,clip,keepaspectratio]{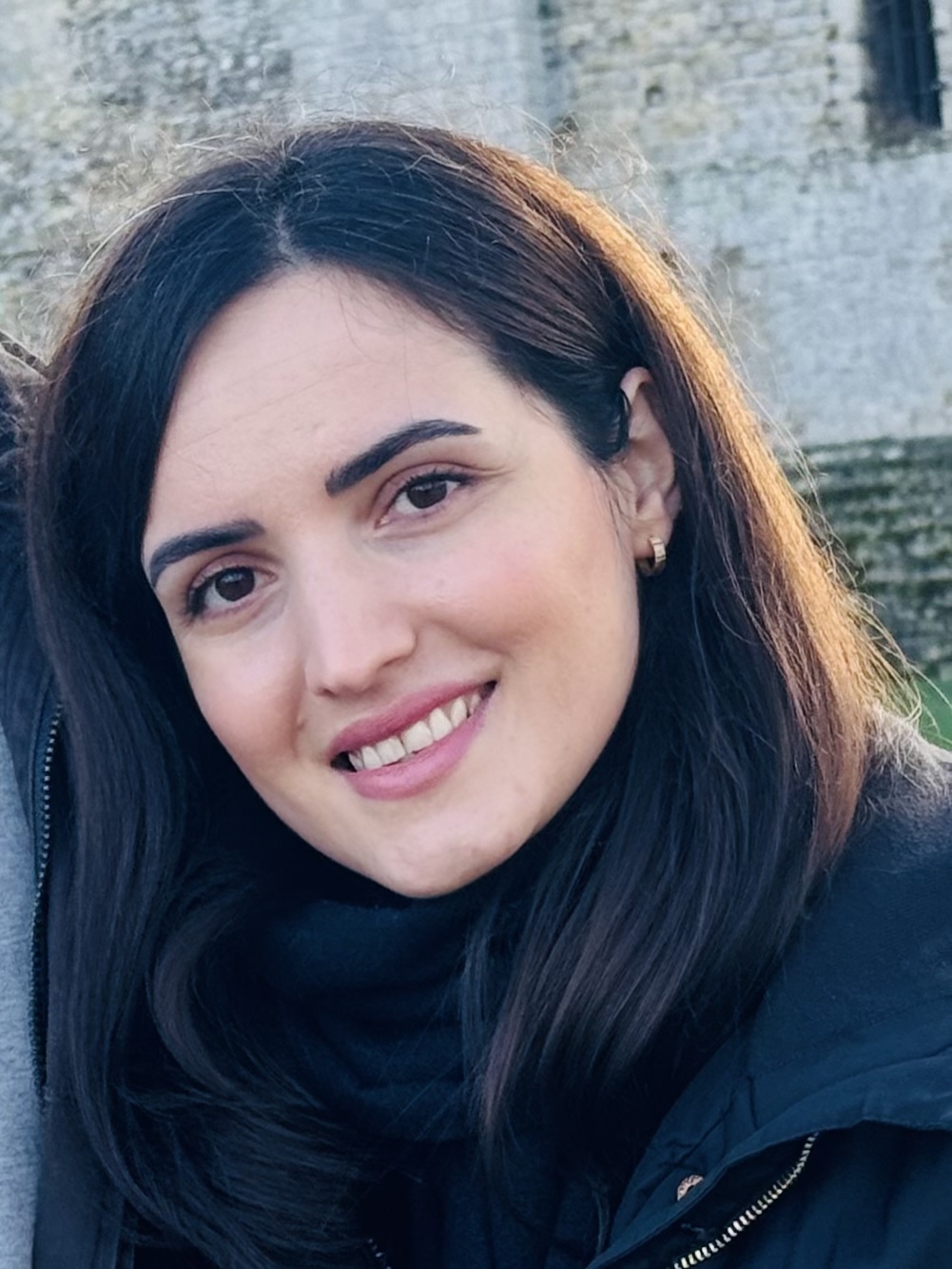} 
\end{minipage}%
\hfill%
\begin{minipage}{0.85\textwidth}
\textbf{Hadiseh~Rezaei} received her PhD in Computer Science (Cybersecurity and Machine Learning) from the University of Portsmouth, UK. Her research spans federated learning-based intrusion detection, adversarial machine learning, and privacy-preserving training, with publications in high-impact IEEE, Springer, and Elsevier journals. She is also an active peer reviewer for leading journals and has industry experience delivering security risk and compliance programmes aligned with ISO/IEC~27001 and NIST frameworks.
 \end{minipage}%

\hfill \break

\noindent\begin{minipage}{0.14\textwidth}
\includegraphics[width=1.05in,height=1.05in,clip,keepaspectratio]{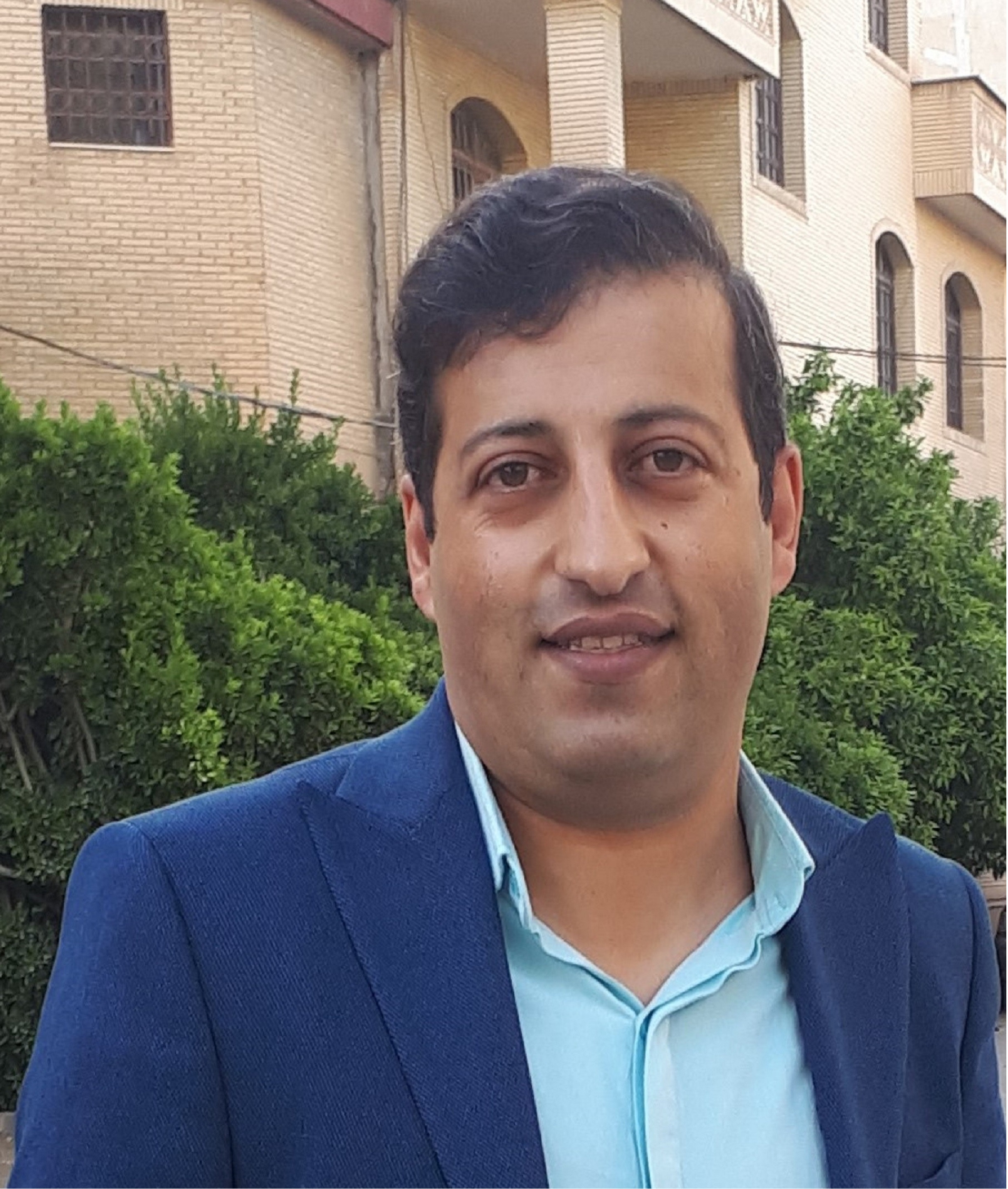} 
\end{minipage}%
\hfill%
\begin{minipage}{0.85\textwidth}
\textbf{Rahim~Taheri} earned his B.Sc. in Computer Engineering from Bahonar Technical College of Shiraz in 2007. Subsequently, he completed his M.Sc. and Ph.D. in Computer Networks at Shiraz University of Technology in 2015 and 2020, respectively. Currently, he serves as a Lecturer in Cyber Security and Forensics at the University of Portsmouth, UK. Prior to joining the University of Portsmouth, he worked as a post-doctoral research associate at King’s Communications, Learning, and Information Processing (Kclip) Lab, King’s College London, UK. His primary research interests encompass computer networks, machine learning applications in security, adversarial machine learning, and federated learning.
\end{minipage}%

\hfill \break

\noindent\begin{minipage}{0.14\textwidth}
\includegraphics[width=0.93in,height=1.2in,clip,keepaspectratio]{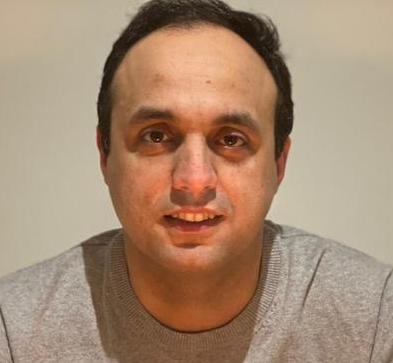} 
\end{minipage}%
\hfill%
\begin{minipage}{0.85\textwidth}
\textbf{Mohammad~Shojafar} received a Ph.D. degree in ICT from the Sapienza University of Rome, Rome, Italy, in 2016, with an ``Excellent'' degree. He is a Senior Lecturer (Associate Professor) in network Security, an Intel Innovator, a Professional ACM Member and ACM Distinguished Speaker, a Fellow of the Higher Education Academy, and a Marie Curie Alumni, working with the 5G \& 6G Innovation Centre (5G/6GIC), Institute for Communication Systems, University of Surrey, U.K. Before joining 5G/6GIC, he was a Senior Researcher and a Marie Curie Fellow in the SPRITZ Security and Privacy Research Group with the University of Padua, Italy. He secured around $\pounds$1.9M as PI in various EU/U.K. projects. For additional information: \url{http://mshojafar.com} 
\end{minipage}%

\hfill \break

\end{document}

\section{\textbf{Simulation and Evaluation}
\item }\label{Simulation and Evaluation}
\subsection{}

\textcolor{red}{After this need to add new texts}

\subsection{Proposed Validation Indices for Automatic Clustering}\label{sec:3.2}

\subsection{Time Complexity}\label{sec:3.3}

\section{Experimental Evaluation}\label{resultAnalysis}	

\subsubsection{Optimization Algorithm Settings}\label{OptimizationAlgorithm}

\subsubsection{Test metrics}\label{Testmetrics}

\subsubsection{Fitness Function}\label{FitnessFunction} 

\subsubsection{Schematic presentations of samples}\label{schematic}

\subsubsection{Evaluation of fitness function among AI methods}

\subsubsection{Evaluation of average cluster differences}

\subsubsection{Evaluation frequencies among AI methods}

\subsection{Result Discussions}

\section{Conclusions and Future Direction} \label{conclusion}  
\bibliographystyle{plain}
\bibliography{ref}

nces \bib{}liographystyle{IEEEtran}

\end{document}